\documentclass[twocolumn,aps,prd,preprintnumbers,showpacs,superscriptaddress,nofootinbib,amsmath,amssymb,floats,floatfix,showkeys,notitlepage]{revtex4}

\usepackage{lipsum}
\usepackage{graphicx}
\usepackage{subfigure}
\usepackage{palatino}
\usepackage{changes}
\usepackage{hyperref}
\hypersetup{colorlinks=true,linkcolor=blue,urlcolor=blue,citecolor=blue}
\usepackage[toc,page]{appendix}
\usepackage[normalem]{ulem}
\usepackage{adjustbox}
\usepackage{latexsym}
\usepackage{amsmath}
\usepackage{amssymb}
\usepackage{amsfonts}
\usepackage{times}
\usepackage{dcolumn}
\usepackage{bm}
\usepackage{tikz}
\usepackage{bigints}
\usepackage{array,tabularx,multirow}
\usepackage[tracking=true]{microtype}
\SetTracking{}{500}
\SetTracking{encoding={*}, shape=sc}{40}
\UseRawInputEncoding 
\allowdisplaybreaks

\begin{document} \sloppy

\title{Generalized Extended Uncertainty Principle Black Holes: Shadow and lensing in the macro- and microscopic realms}

\author{Nikko John Leo S. Lobos}
\email{njllobos.mp@tip.edu.ph}
\affiliation{Mathematics and Physics Department, Technological Institute of the Philippines, 363 Pascual Casal St, Quiapo, Manila, 1001 Metro Manila}
\affiliation{Physics Department, De La Salle University, 2401 Taft Avenue, Manila, 1004 Philippines}

\author{Reggie C. Pantig}
\email{reggie.pantig@dlsu.edu.ph}
\affiliation{Physics Department, De La Salle University, 2401 Taft Avenue, Manila, 1004 Philippines}


\begin{abstract}
Motivated by the recent work about the Extended Uncertainty Principle (EUP) black holes \cite{Mureika:2018gxl}, we present in this study its extension called the Generalized Extended Uncertainty Principle (GEUP) black holes. In particular, we investigated the GEUP effects on astrophysical and micro-black holes. First, we derive the expression for the shadow radius to investigate its behavior as perceived by a static observer located near and far from the black hole. Constraints to the large fundamental length scale $L*$ up to $2\sigma$ level were also found using the EHT data: for Sgr. A*, $L* = 5.716\text{x}10^{10}$ m, while for M87*, $L* = 3.264\text{x}10^{13}$ m. Under the GEUP effect, the value of the shadow radius behaves the same way as the Schwarzschild case due to a static observer, and the effect only emerges if the mass of the black hole $M$ is around the order of magnitude of $L_*$ (or $l_\text{Pl}$). In addition, the GEUP effect increases the shadow radius for astrophysical black holes, but the reverse happens for micro-black holes. We also explored GEUP effects to the weak and strong deflection angles as an alternative analysis. For both realms, a time-like particle gives a higher value for the weak deflection angle. Similar to the shadow, the deviation is seen when the values of $L_*$ and $M$ are close. The strong deflection angle gives more sensitivity to GEUP deviation at smaller masses in the astrophysical scenario. However, the weak deflection angle is a better probe in the micro world.
\end{abstract}



\maketitle

\section{Introduction} \label{intro}
Black hole theory has never been more exciting than before when the Event Horizon Telescope Collaboration (EHT) revealed the first image of a black hole in M87 galaxy \cite{EventHorizonTelescope:2019dse}, and more recently, the black hole Sgr. A* in our galaxy \cite{EventHorizonTelescope:2022xnr}. These pictures, with very special algorithms, provided further evidence that black holes exist in Nature. Black holes are compact objects with gravity so strong that not even light can escape its gravitational grip.

Black hole solutions are found by solving the Einstein field equation, and the simplest black hole model that is static and spherically symmetric was found by Karl Schwarzschild \cite{Schwarzschild_1916}. Later on, the metric of a spinning black hole, which is static and axisymmetric, was found by Roy Kerr \cite{Kerr:1963ud}. Conceptually, black holes are massive objects where all the mass is concentrated into a point, thus giving the object an infinite density. In essence, it is no doubt that there must be some interplay between gravity and quantum mechanics in these extreme regions. Indeed, black holes are laboratories where one can probe the quantum nature of gravity \cite{rovelli_2004}.

Central to the microscopic realm is the Heisenberg Uncertainty Principle (HUP), which states that
\begin{equation} \label{ehup}
    \Delta x \Delta p \geq \frac{\hbar}{2},
\end{equation}
which is derived from the commutation relation of the position and momentum operators $\hat{x}$ and $\hat{p}$. That is, $[\hat{x},\hat{p}]=i\hbar$. Eq. \eqref{ehup} can provide limitations in testing predictions, but in principle, a hypothetical energy probe can still detect very short distance scales. The main problem is that beyond the Planck length $l_\text{Pl}$, there is no guarantee that the spacetime observed is still smooth. Such a chaotic spacetime in the microscopic realm is called the quantum foam \cite{PhysRev.97.511}. It is only then that the HUP must be modified to accommodate the Planck length, and the most accepted modification is called the Generalized Uncertainty Principle (GUP) \cite{PhysRevD.49.5182,Maggiore_1993_1,Maggiore_1993_2}, which adds a quadratic uncertainty in momentum:
\begin{equation} \label{egup}
    \Delta x \Delta p \geq 1+ \beta l_\text{Pl}^2 \Delta p^2.
\end{equation}
Here, $\beta$ is a dimensionless quantity usually taken as unity and can be either positive or negative \cite{AMATI198941}.

As Nature is fond of symmetry and duality, similar to the yin-yang symbol, it is only natural to suspect that if there is a minimum fundamental length, there must be a large fundamental length scale in our Universe. Hence, the GUP is naturally extended in Ref. \cite{Bambi_2008}, to include the large fundamental length $L_*$ through a quadratic correction in the position uncertainty. That is,
\begin{equation} \label{eEup}
    \Delta x \Delta p \geq 1 +\alpha\frac{\Delta x^2}{L_*^2},
\end{equation}
which is commonly called the Extended Uncertainty Principle (EUP). Here, $\alpha$ is also a dimensionless constant. It was also derived from first principles \cite{Costafilho_2016}. While GUP is commonly analyzed in the literature due to its vast application in the microscopic world \cite{Tawfik2014}, the application of EUP seems to be dearth in the literature. For instance, the analysis of EUP effects on the thermodynamics of FRW Universe \cite{Zhu_2009} was analyzed long ago and a year later applied to the geometry of dS and AdS spacetime \cite{Mignemi_2010}. The effects of the EUP correction has also been studied in Rindler and cosmological horizons \cite{Dabrowski:2019wjk}, relativistic Coulomb potential \cite{Hamil:2020jns}, bound-state solutions of the two-dimensional Dirac equation with Aharonov–Bohm-Coulomb interaction \cite{Hamil:2020xud}, J\"uttner gas \cite{Moradpour:2021ymp}. With the help of the GUP and EUP parameters, bounds for the Hubble parameter's value were also studied to resolve the Hubble tension \cite{Aghababaei:2021gxe}. It is only recently that EUP correction has been applied in the context of black holes \cite{Mureika:2018gxl}, where $r_{h} \sim \Delta x$ as the gravitons are the quantum particles inside such confinement. Since then, various authors have explored the black hole with EUP correction (see Refs. \cite{Lu:2019wfi,Kumaran:2019qqp,Cheng:2019zgc,Hassanabadi:2021kyv,Hamil:2021ilv,Okcu:2022sio,Pantig:2021zqe,Hamil:2022mca,Chen:2022ngd}).

We are motivated to continue the analysis made in Ref. \cite{Mureika:2018gxl}, where we further investigate the most general form of the uncertainty principle \cite{Kempf:1994su}
\begin{equation} \label{egeup}
    \Delta x \Delta p \geq 1+ \beta l_\text{Pl}^2 \Delta p^2+\alpha\frac{\Delta x^2}{L_*^2}
\end{equation}
as applied to the shadow cast and gravitational lensing of astrophysical black holes and micro-black holes \cite{Calmet:2014dea,Hawking:1971ei}. To this aim, the black hole metric that contains the GEUP correction must be expressed as \cite{Mureika:2018gxl}
\begin{equation} \label{eqn:1}
    ds^{2}= - A(r) dt^{2}+ B(r) dr^{2}+ C(r) d\theta^{2}+ D(r) d\varphi^{2} ,
\end{equation}
where
\begin{align} \label{eqn:2}
    A(r) &= 1 - \frac{2\mathcal{M}}{r}, \quad B(r) = A(r)^{-1}, \nonumber \\ 
    C(r) &= r^2, \quad D(r) = r^2 \sin ^{2} \theta.
\end{align}
With the GEUP correction in Eq. \eqref{egeup} the mass of the black hole $M$ is now corrected as\cite{Mureika:2018gxl},
\begin{equation} \label{eqn:3}
    \mathcal{M} = M\left(1 + \frac{4\alpha M^2}{L^2_*} + \frac{\beta \hbar}{2 M^2} \right).
\end{equation}
Here, we initially show $\hbar$ to emphasize the quantum correction for quantum particles. Note that since $M$ is geometrized, we can relate $\hbar$ to the Planck length representing the known minimal length $l_\text{Pl} = 1.616\text{x}10^{-35}\text{ m}$. Furthermore, $\alpha = \beta = 1$, and $L_*$'s value will be estimated based on the observational constraints from the EHT in Sect. \ref{sec2}. First, we will explore the behavior of the shadow radius of the object being considered (i.e. SMBH for macroscopic and some elementary particles for the microscopic realm). Shadows are important since they can reveal imprints that allow one to test gravity theories in the strong field regime. Shadows are first studied by \cite{Synge1966}. In 1979, Luminet gave the formula for the angular radius of the shadow \cite{Luminet1979}. Then several authors have explored the shadows of quantum black holes \cite{Konoplya:2019xmn,Hu:2020usx,Tamburini:2021inp,Devi:2021ctm,Anacleto:2021qoe,Xu:2021xgw,Karmakar:2022idu,Rayimbaev:2022hca}.
In this paper, we are also interested in probing the GEUP effects using the strong and weak deflection angles. Gravitational lensing is one of the most successful tools as it verified Einstein's general theory of relativity \cite{Dyson:1920cwa} in 1919 through the Sun's solar eclipse. Since then, it has been crucial in probing various tests of gravitation theories. Several tools have been developed \cite{Virbhadra:1999nm,Bozza:2001xd,Bozza:2002zj}, and in 2008, the Gauss-Bonnet theorem on the optical geometries in asymptotically flat spacetimes was developed \cite{Gibbons:2008rj}. It was extended by Werner \cite{Werner_2012} to include stationary spacetimes in the Finsler-Randers type optical geometry on Nazim's osculating Riemannian manifolds. Ishihara and others then found a way to extend the GBT to incorporate finite distance effects \cite{Ishihara:2016vdc,Ishihara:2016sfv}, which also applies to non-asymptotic spacetimes. Finally, instead of using points at infinity as integration domain for the GBT, \cite{Li:2020wvn} used the photonsphere to naturally find an alternative to the Ishihara method, which also accommodates the deflection angle of massive particles. For recent works about quantum black holes’ deflection angles, see Refs. \cite{Pantig:2021zqe,Kumaran:2019qqp,Xu:2018wgi,Zhang:2021yyi,Fu:2021fxn,Lu:2021htd,Jusufi:2018kmk}. 

We organized the paper as follows: We devote Sect. \ref{sec2} in exploring the shadow behavior of the GEUP black hole and microscopic entities if they are viewed as quantum black holes. In Sect. \ref{sec3}, we used the Gauss-Bonnet theorem to study the weak deflection angle of the mentioned objects. Sect. \ref{sec4} is about the strong deflection angle, which is a generalization of the weak deflection angle in the previous section. Then, in Sect. \ref{conclu}, we formulate our conclusion based on the results of the prior sections. In this paper also, we used geometrized units wherein $G = c = 1$, and the metric signature ($-,+,+,+$). Hence, we can replace $\hbar$ in Eq. \eqref{eqn:3} with the Planck length.

\section{Shadow and constraints to the large fundamental length scale} \label{sec2}
In this section, we will study the shadow of the GEUP black hole. Due to $r$ and $t$ independence of the metric, such symmetry allows us to analyze light-like geodesics along the equatorial plane ($\theta = \pi/2$) without compromising generality. Thus, $D(r) = C(r)$ in the metric Eq. \eqref{eqn:1}. These geodesics can be derived through the Lagrangian
\begin{equation}
    \mathcal{L} = \frac{1}{2}\left( -A(r) \dot{t} +B(r) \dot{r} + C(r) \dot{\phi} \right).
\end{equation}
Through the variational principle, the Euler-Lagrange equation gives two constants of motion
\begin{equation} \label{econs}
    E = A(r)\frac{dt}{d\lambda}, \quad L = C(r)\frac{d\phi}{d\lambda},
\end{equation}
where we can define the impact parameter as
\begin{equation}
    b \equiv \frac{L}{E} = \frac{C(r)}{A(r)}\frac{d\phi}{dt}.
\end{equation}
For light-like geodesics, the metric can be set as $ds^2 = 0$, and using Eq. \eqref{econs}, we can obtain the orbit equation as
\begin{equation}
    \left(\frac{dr}{d\phi}\right)^2 =\frac{C(r)}{B(r)}\left(\frac{h(r)^2}{b^2}-1\right),
\end{equation}
where, by definition \cite{Perlick2015},
\begin{equation}
    h(r)^2 = \frac{C(r)}{A(r)}.
\end{equation}
Through the above equation, we can obtain the location of the photonsphere by taking $h'(r) = 0$, where prime denotes taking the derivative with respect to $r$. To this end, since the mass $M$ is just imbued with quantum correction, the location of the photonsphere is simply
\begin{equation} \label{erph}
    r_\text{ph} = 3\mathcal{M}.
\end{equation}

\begin{figure*}
    \centering
    \includegraphics[width=0.49\textwidth]{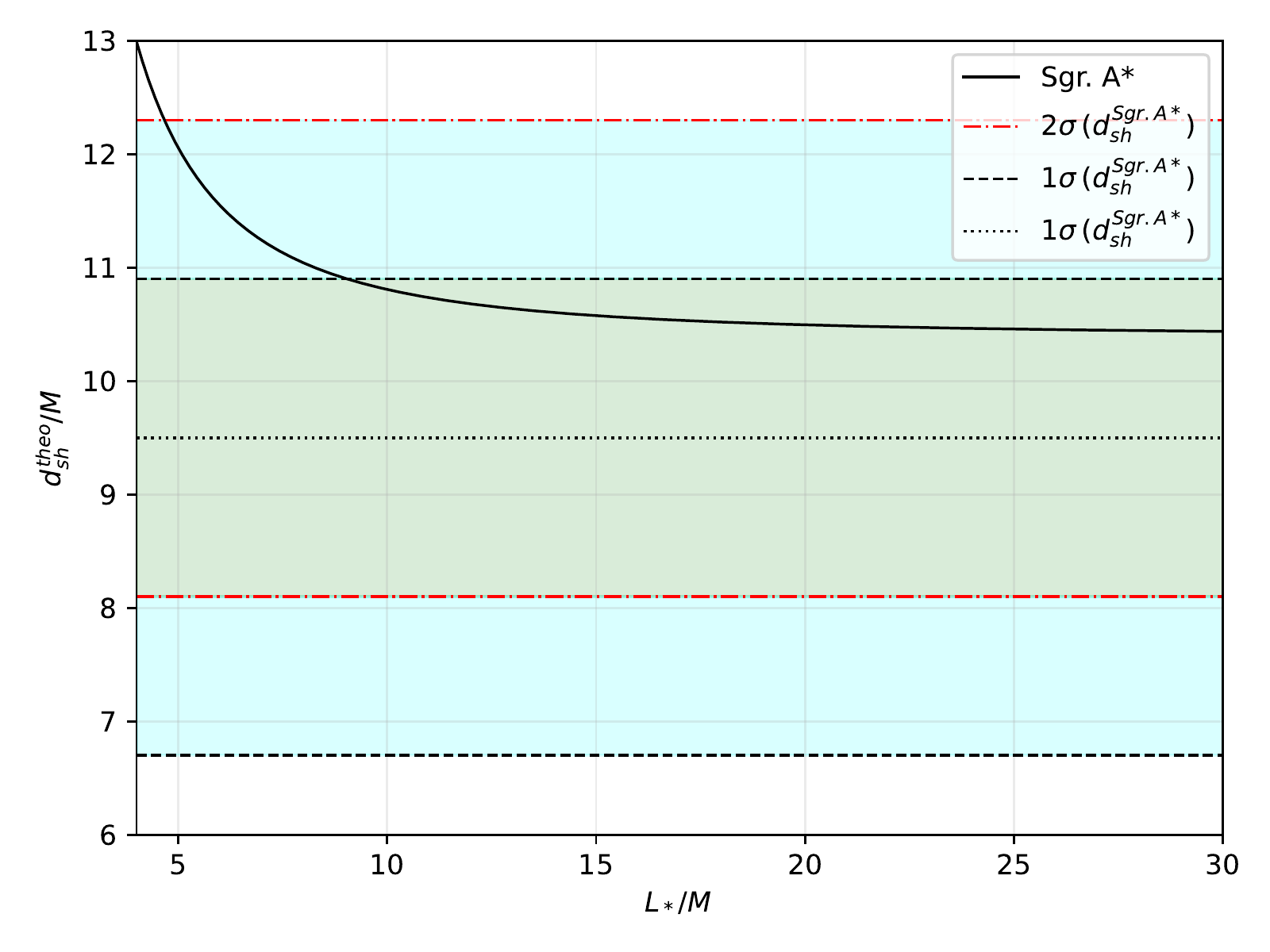}
    \includegraphics[width=0.49\textwidth]{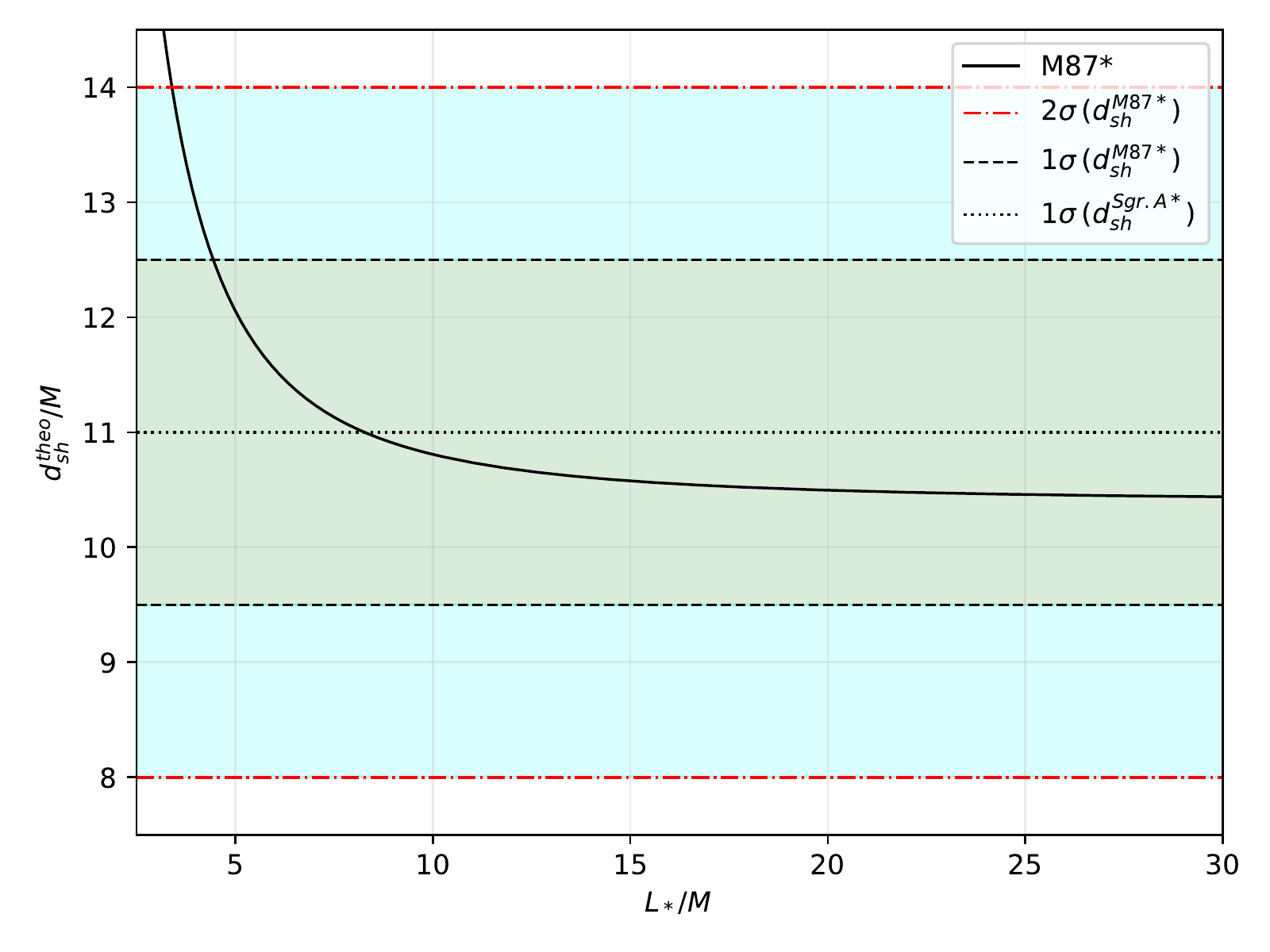}
    \caption{Observational contraints for $L_*$ for Sgr. A* and M87*. Left: 1$\sigma$ - $L_* \sim 5.716\text{x}10^{10}$ m, 2$\sigma$ - $L_* \sim 2.985\text{x}10^{10}$ m. Right: 1$\sigma$ - $L_*\sim4.224\text{x}10^{13}$ m, 2$\sigma$ - $L_*\sim3.264\text{x}10^{13}$ m. At the mean, $L_*\sim7.950\text{x}10^{13}$ m.}
    \label{shacons}
\end{figure*}

Our concern in this section is how the observer will perceive the GEUP black hole at near and far away locations. Let the observer be at the coordinates $(t_\text{obs},r_\text{obs},\theta_\text{obs} = \pi/2, \phi_\text{obs} = 0)$. Then, with simple geometry, the observer can construct \cite{Perlick:2018} the relation
\begin{equation}
    \tan(\alpha_{\text{sh}}) = \lim_{\Delta x \to 0}\frac{\Delta y}{\Delta x} = \left(\frac{C(r)}{B(r)}\right)^{1/2} \frac{d\phi}{dr} \bigg|_{r=r_\text{obs}},
\end{equation}
which can be written in another way as
\begin{equation} \label{eangrad}
    \sin^{2}(\alpha_\text{sh}) = \frac{b_\text{crit}^{2}}{h(r_\text{obs})^{2}},
\end{equation}
where $b_\text{crit}$ is a function of the photonsphere given in Eq. \eqref{erph}. A spacetime may have a different expression for $h(r)$, thus for a general spacetime, the critical impact parameter can be obtained as \cite{Pantig:2022ely}
\begin{align} \label{ebcrit}
    &b_\text{crit}^2 = \frac{h(r_\text{ph})}{\left[B'(r_\text{ph})C(r_\text{ph})-B(r_\text{ph})C'(r_\text{ph})\right]} \Bigg[h(r_\text{ph})B'(r_\text{ph})C(r_\text{ph}) \nonumber \\
    &-h(r_\text{ph})B(r_\text{ph})C'(r_\text{ph}) 
    -2h'(r_\text{ph})B(r_\text{ph})C(r_\text{ph}) \Bigg],
\end{align}
which, for the GEUP black hole, we find
\begin{equation} \label{ebcrit2}
    b_\text{crit}^2 = 27\mathcal{M}^2.
\end{equation}
Finally, we can now obtain the behavior of the shadow radius, applicable for both macroscopic and quantum black holes as
\begin{equation} \label{eRsh}
    R_\text{sh} = 3\mathcal{M}\sqrt{3\left(1-\frac{2\mathcal{M}}{r_\text{obs}}\right)}.
\end{equation}
Note that the above expression is valid even when the static observer is near the black hole. In addition, if $r_\text{obs} \to \infty$, we can approximate Eq. \eqref{eRsh} simply as $R_\text{sh} = 3\sqrt{3}\mathcal{M}$. Let us start first with astrophysical black holes, such as Sgr. A* and M87*, and discuss first some observational constraints. According to papers in Refs. \cite{EventHorizonTelescope:2019dse,EventHorizonTelescope:2022xnr}, the mass, distance from Earth, and angular shadow diameter of M87* are \cite{PhysRevD.95.104011} $M_\text{M87*} = 6.5 \pm 0.90$x$10^9 \: M_\odot$, $D = 16.8$ Mpc, and $\alpha_\text{M87*} = 42 \pm 3 \:\mu$as respectively. For Sgr. A*, these are $M_\text{Sgr. A*} = 4.3 \pm 0.013$x$10^6 \: M_\odot$ (VLTI), $D = 8277\pm33$ pc, and $\alpha_\text{Sgr. A*} = 48.7 \pm 7 \:\mu$as (EHT) respectively. The diameter of the shadow size using these empirical data and in units of the black hole mass can be calculated using
\begin{equation} \label{edia}
    d_\text{sh} = \frac{D \theta}{M}.
\end{equation}
Then we have the diameter of the shadow image of M87* and Sgr. A* as $d^\text{M87*}_\text{sh} = (11 \pm 1.5)M$, and $d^\text{Sgr. A*}_\text{sh} = (9.5 \pm 1.4)M$ respectively.
Meanwhile, the theoretical shadow diameter can be obtained via $d_\text{sh}^\text{theo} = 2R_\text{sh}$. The observational constraints' results are plotted in Fig. \ref{shacons}.

Theoretically, let us now consider how the static observer perceives the shadow radius at different locations in the radial coordinate for different values of $L_*$. In the literature, only the situation $r_\text{obs} \to \infty$ were considered \cite{Mureika:2018gxl,Pantig:2021zqe}.
\begin{figure*}
    \centering
    \includegraphics[width=0.49\textwidth]{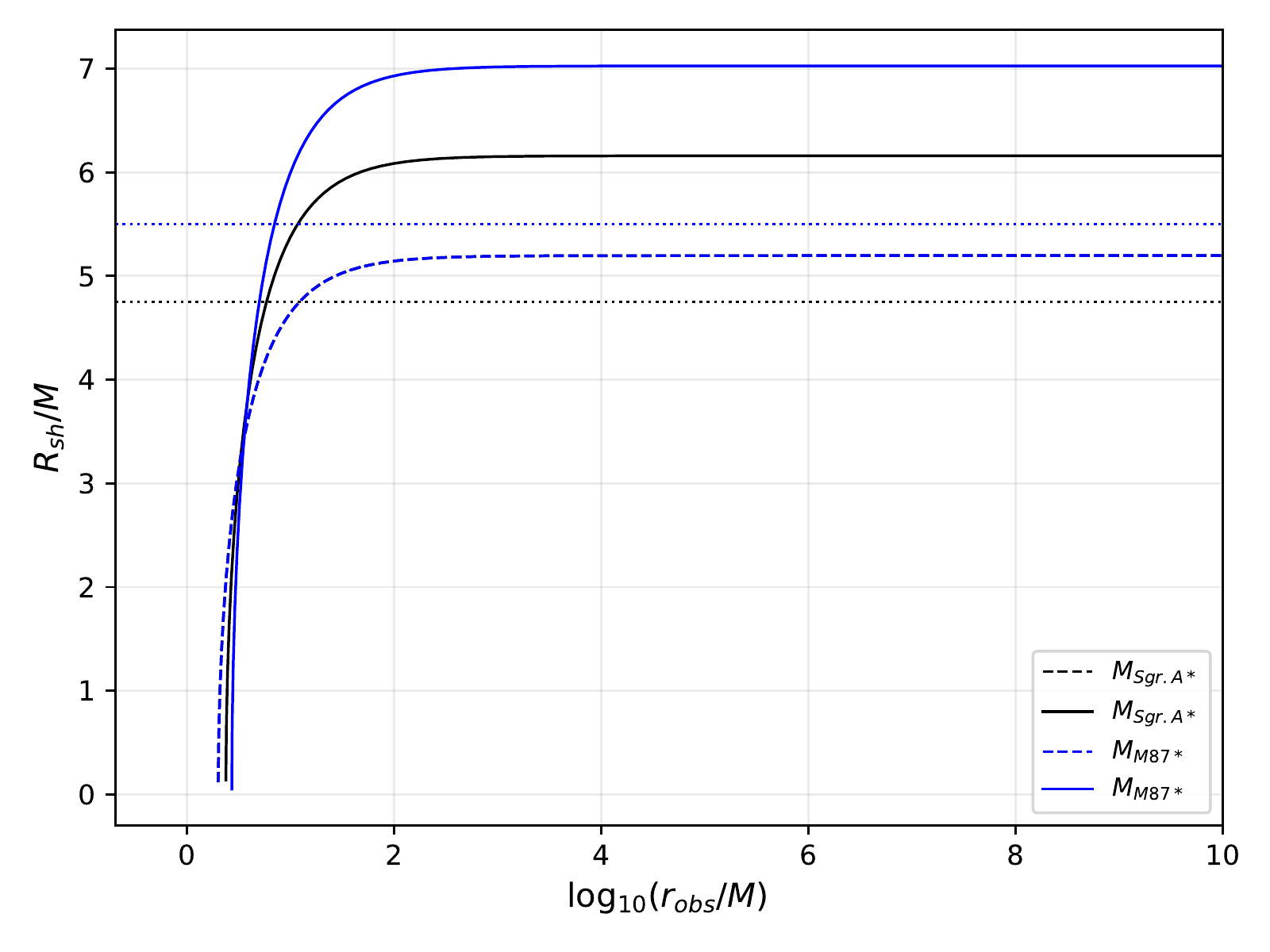}
    \includegraphics[width=0.49\textwidth]{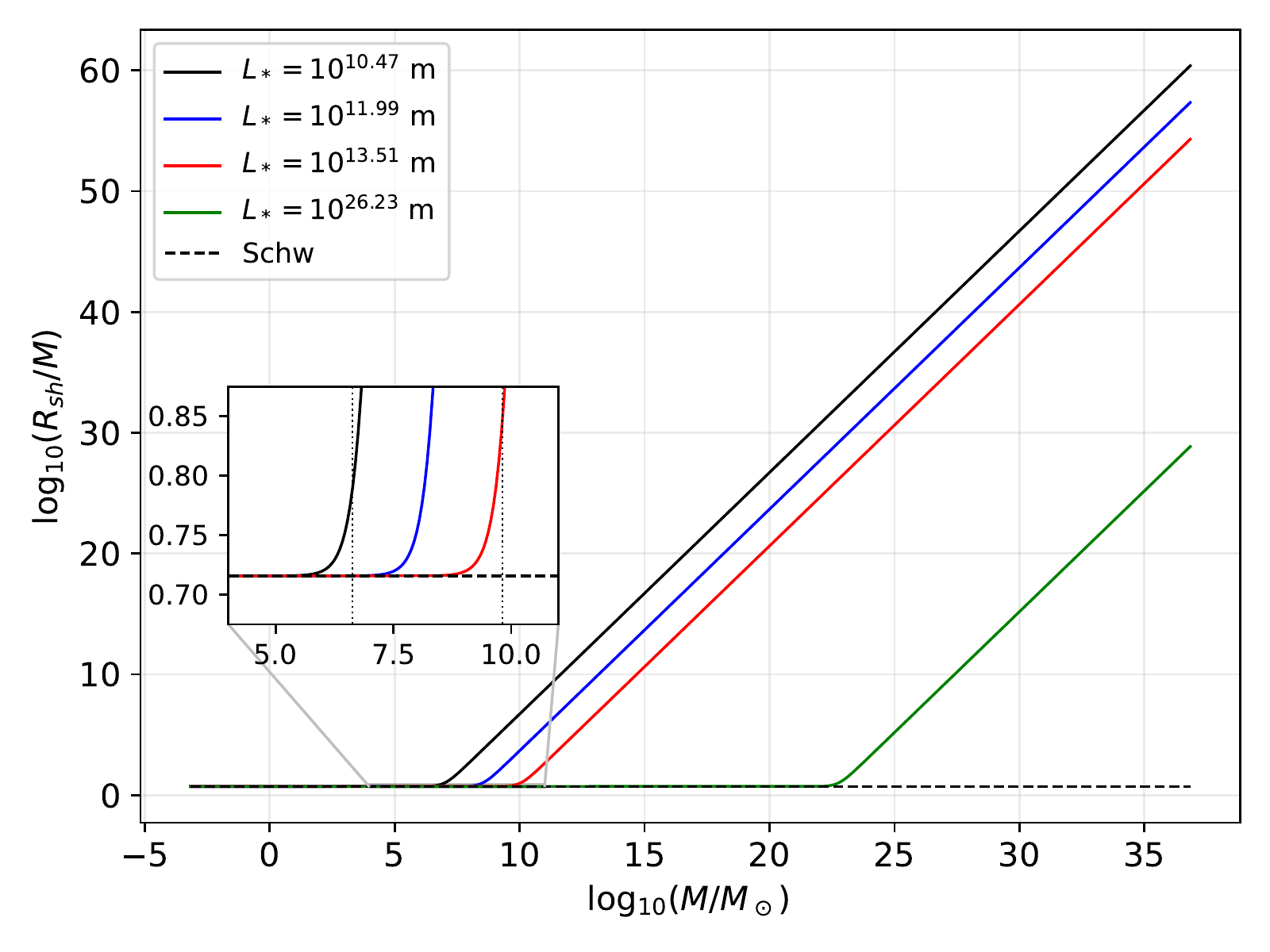}
    \caption{Left: Plot of the shadow radius of Sgr. A* and M87* with observer location dependency. The dashed line represents the Schwarzschild case and the solid line for the GEUP case. The horizontal black and blue dotted lines represent the shadow radius of Sgr. A* and M87* based on EHT data. Right: How the shadow radius behave for different black hole masses. The black and blue vertical line in the inset plot represents the mass of the Sgr. A* and M87*, respectively.}
    \label{shasmbh}
\end{figure*}
In the left plot, the dashed line is the Schwarzschild case for both SMBHs, which overlaps. The shadow radius coming from empirical data is also shown for comparison. Notice that the GEUP effect merely increases the shadow radius while the curve's trend is the same as in the Schwarzschild case.  In the plot to the right, we see how the shadow radius behaves due to the GEUP effect. For instance, deviations begin to manifest if the value of $L_*$ is close to the mass of the black hole, which is also clearly seen by the green line as we include $L_*$ that is comparable to the Hubble length. In this scenario, the effect of the parameters in the microscopic realm does not even manifest.

\begin{figure*}
    \centering
    \includegraphics[width=0.49\textwidth]{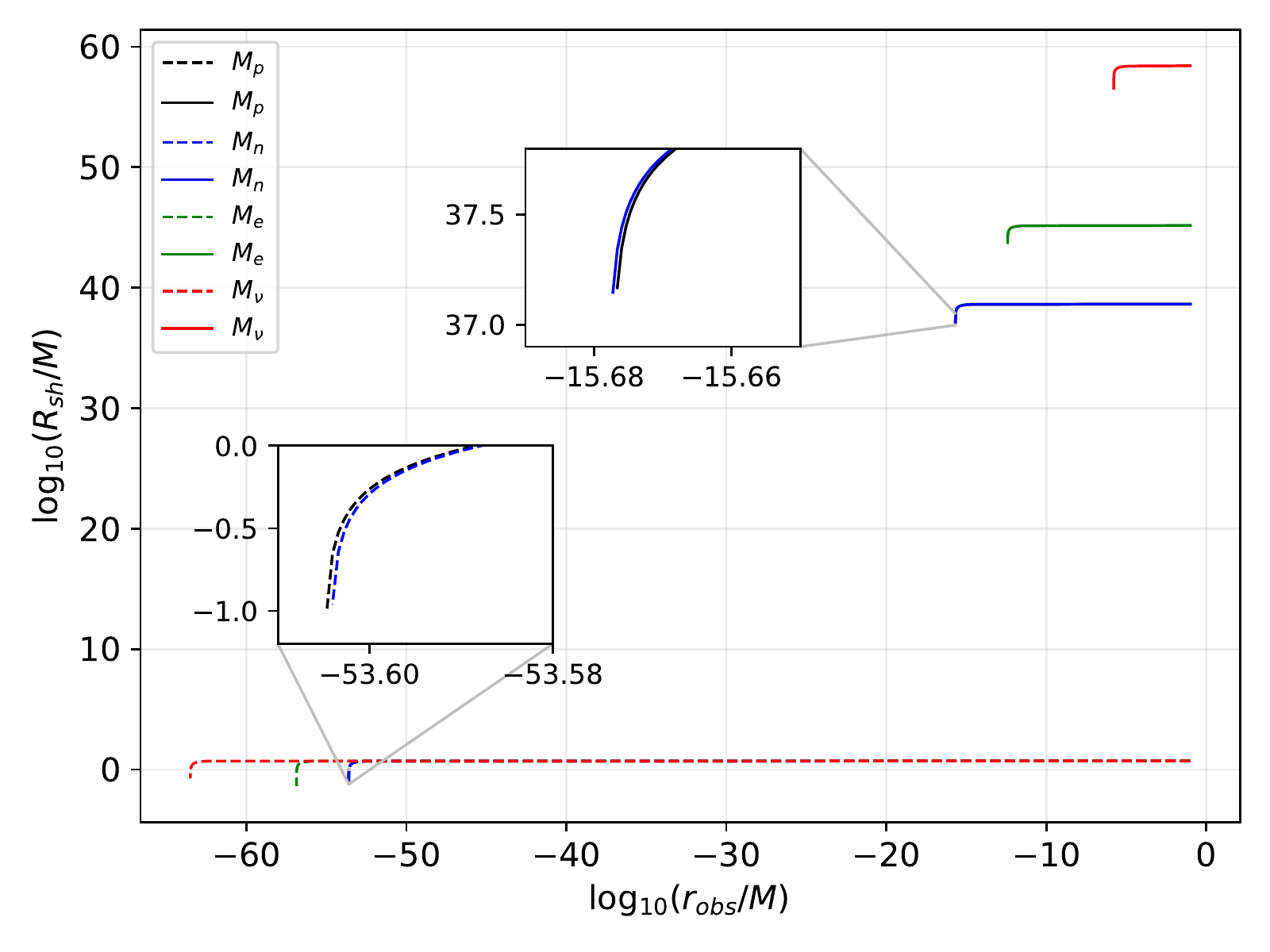}
    \includegraphics[width=0.49\textwidth]{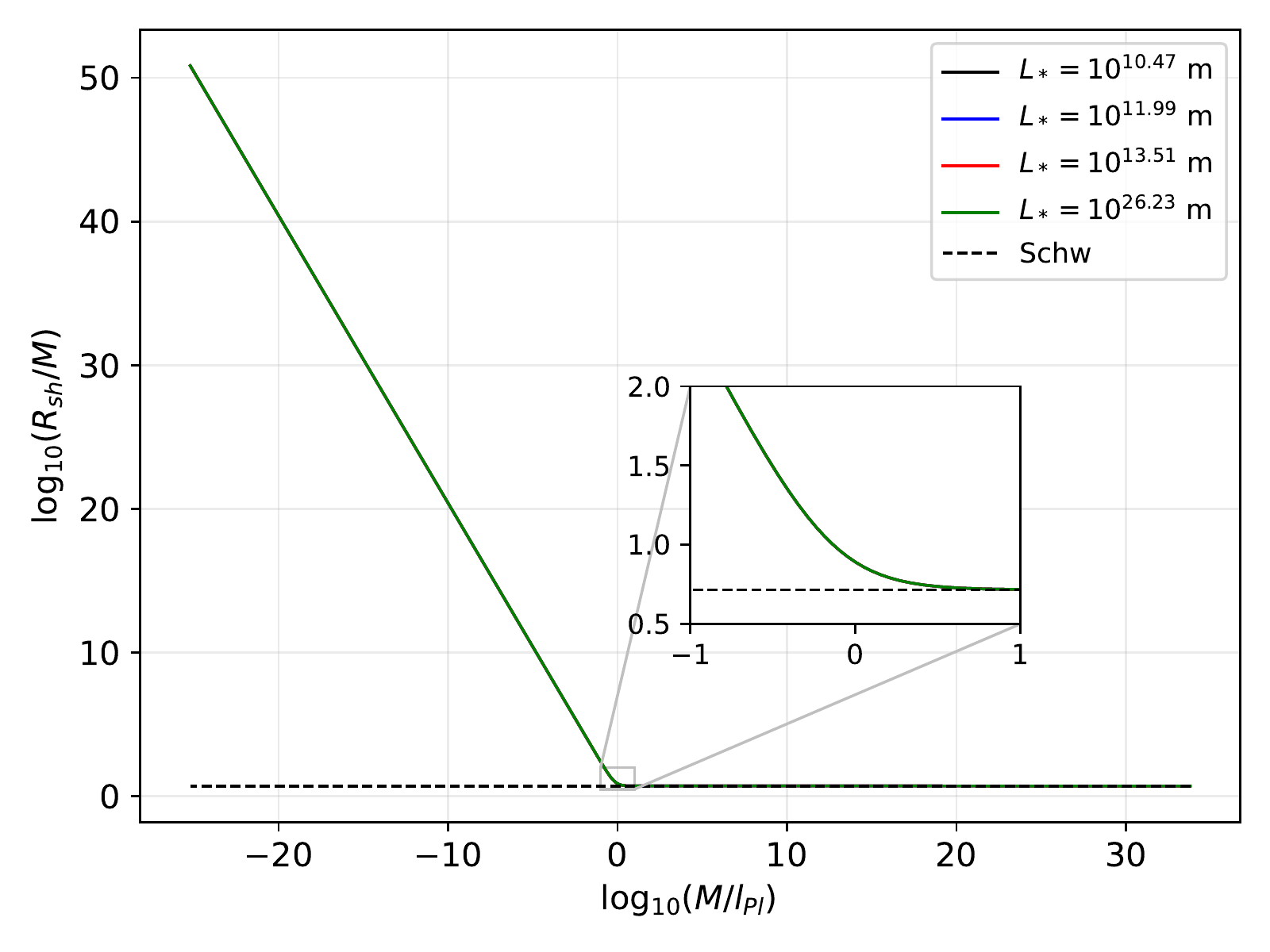}
    \caption{Left: Observer dependent shadow radius of some elementary particles. Right: The shadow radius plotted under the assumption that the observer/detector is at $r_\text{obs}>>m$ for different values of $L_*$. The overlapping of these lines mean that $L_*$ has no effect in the microscopic realm.}
    \label{shaelem}
\end{figure*}

Eq. \eqref{eRsh} also admits analysis for quantum black holes. We plotted our results in Fig. \ref{shaelem}.
In the left figure, we consider the case wherein the static observer may be represented by a detector that can probe masses as small as the proton, neutron, electron, and neutrino \cite{Eitel:2005hg}, where we have used their geometrized masses. The dashed and solid lines represent the Schwarzschild and GEUP cases, respectively. The figure to the right reveals that $L_*$ is indeed irrelevant in the microscopic realm. Nonetheless, with the GUP correction, the plot reveals the detector's position where the shadow of the particle manifests. Take, for example, the neutrino. Without GUP correction, the shadow radius is around $10^{-63}$ order of magnitude for a wide range of detector's locations. The GUP correction lessens this range and makes the shadow radius larger. For instance, if the detector is at $r = 1.59\text{x}10^{-67}$ m, then the shadow radius is around $R_\text{sh} = 5.03\text{x}10^{-6}$ m. Note how the shadow radius of these particles levels at greater distances. Finally, we observe that without GUP correction, the shadow radius is nearly identical to each other. With the GUP correction, we have seen that as the mass of the particle decreases, the shadow radius tends to increase while the range where a detector can observe it decreases.

\section{Weak deflection angle} \label{sec3}
In this section, we will explore a different phenomenon and examine the effect of the GEUP correction on the weak deflection angle by black holes in the macro- and microscopic realms. To do so, we will exploit the Gauss-Bonnet theorem (GBT). Consider the domain $(D_a,\bar{g})$ that is simply connected over an osculating Riemannian manifold $(\mathcal{M}, \bar{g})$ along some boundaries, and let $\kappa_g$ be the geodesic curvature of $\partial D_a$. Then the Gauss-Bonnet theorem states that \cite{o1992riemannian,Gibbons:2008rj}
\begin{equation}
    \iint_DKdS+\sum\limits_{a=1}^N \int_{\partial D_{a}} \kappa_{g} d\ell+ \sum\limits_{a=1}^N \theta_{a} = 2\pi\chi(D_a),
\end{equation}
where $\chi(D_a)$ is the Euler characteristic, and $\theta_a$ is the exterior angle at the nth vertex.

Although the spacetime herein is asymptotically flat under the GEUP correction, we used the generalized GBT that considers non-asymptotically flat spacetime and massive particle deflection. In Ref. \cite{Li:2020wvn}, the photonsphere $r_\text{ph}$ is the one considered as part of the quadrilateral for integration domain. It is shown that
\begin{equation} \label{eGBTnon}
    \hat{\alpha} = \iint_{_{r_\text{co}}^{R }\square _{r_\text{co}}^{S}}KdS + \phi_{\text{RS}},
\end{equation}
where S and R are the radial positions of the source and receiver, respectively. In this expression, $dS = \sqrt{g}drd\phi$, and $\phi_\text{RS}$ is the coordinate position angle between the source and the receiver defined as $\phi_\text{RS} = \phi_\text{R}-\phi_\text{S}$. Furthermore, $g$ is the determinant of the Jacobi metric in static and spherically symmetric spacetime:
\begin{align} \label{eJac}
    dl^2=g_{ij}dx^{i}dx^{j}
    =(E^2-\mu^2A(r))\left(\frac{B(r)}{A(r)}dr^2+\frac{C(r)}{A(r)}d\Omega^2\right).
\end{align}
Here, $E$ is the energy of the massive particle defined by
\begin{equation} \label{en}
    E = \frac{\mu}{\sqrt{1-v^2}},
\end{equation}
and $v$ is the particle's velocity. As we consider only the equatorial plane due to spherical symmetry, the determinant of the Jacobi metric can be found through
\begin{equation}
    g=\frac{B(r)C(r)}{A(r)^2}(E^2-\mu^2 A(r))^2.
\end{equation}
Following carefully the methodology in Ref. \cite{Li:2020wvn}, we obtained the final expression for the weak deflection angle as
\begin{align} \label{ewda}
    \hat{\alpha} &\sim \frac{\mathcal{M}\left(v^{2}+1\right)}{bv^{2}}\left(\sqrt{1-b^{2}u_\text{R}^{2}}+\sqrt{1-b^{2}u_\text{S}^{2}}\right)
\end{align}
which also involves the finite distance $u_\text{S}$ and $u_\text{R}$. The above expression can still be further approximated as $b^2u^2 \sim 0$:
\begin{align} \label{ewda2}
    \hat{\alpha} \sim \frac{2\mathcal{M}\left(v^{2}+1\right)}{bv^{2}}
\end{align}
For the case of photons where $v=1$, we find
\begin{align} \label{ewda3}
    \hat{\alpha} \sim \frac{4\mathcal{M}}{b}.
\end{align}

\begin{figure*}
    \centering
    \includegraphics[width=0.49\textwidth]{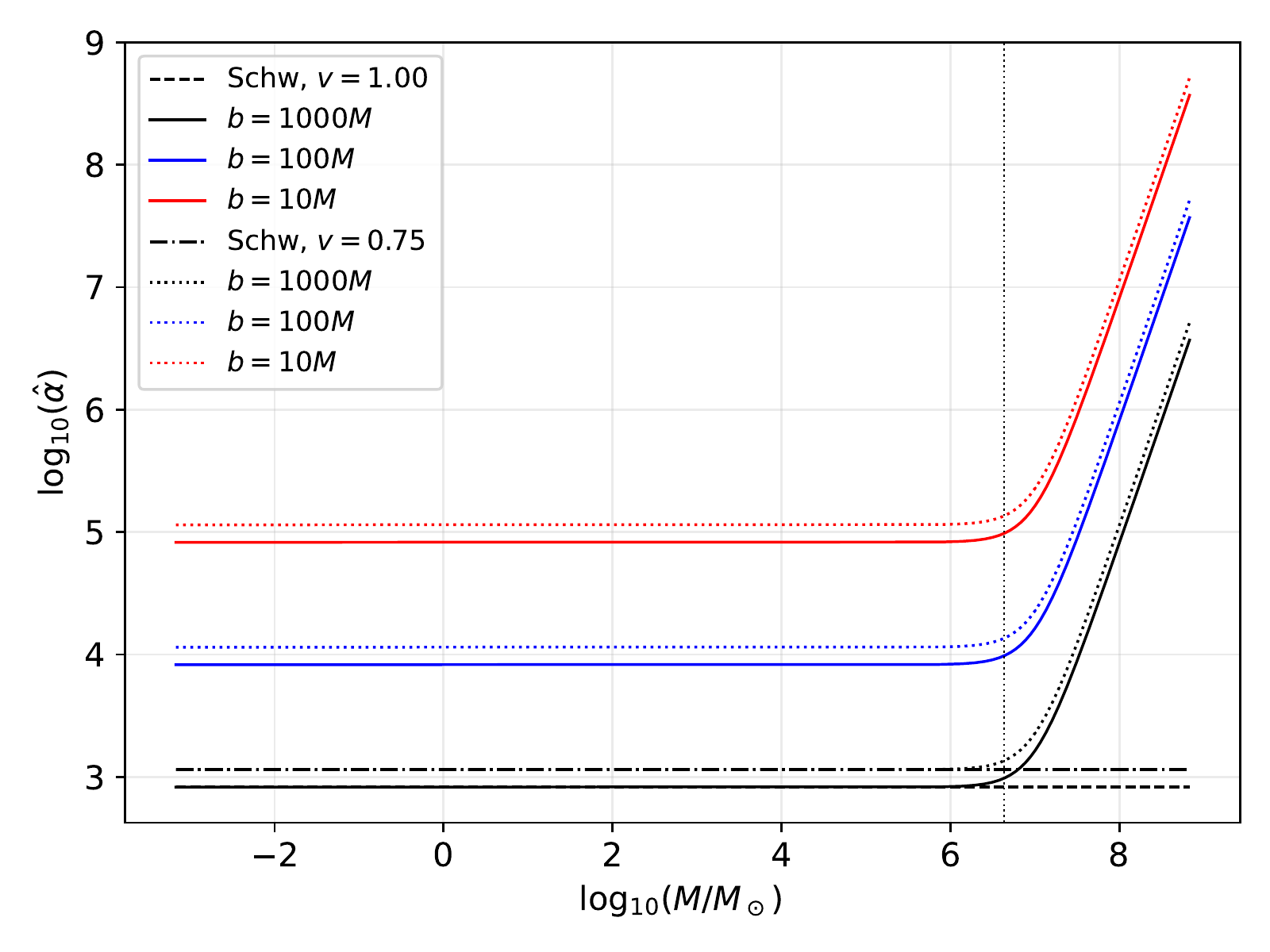}
    \includegraphics[width=0.49\textwidth]{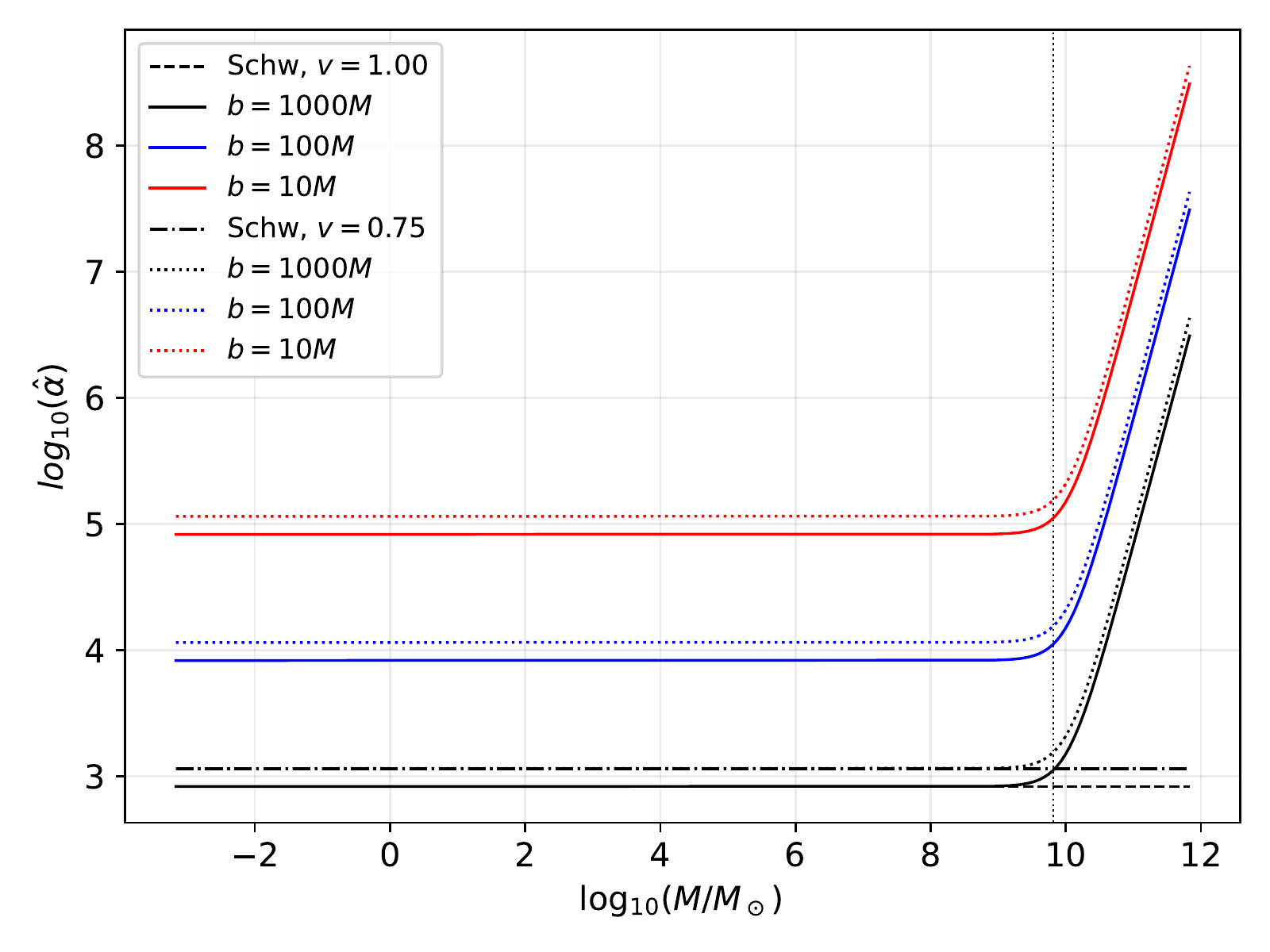}
    \caption{Weak deflection angle by Sgr. A* (left), and M87* (right) for different values of impact parameter. The vertical dotted line is the mass of the black hole considered.}
    \label{wdabh}
\end{figure*}

The weak deflection angle result is usually applied to SMBH. While $\hat{\alpha}$ is usually plotted against the impact parameter $b/M$, in Fig. \ref{wdabh} we are interested in how $\hat{\alpha}$ changes as the BH mass under the effect of GEUP varies. Without the GEUP correction, the plot would only be straight lines. It is observed that similar to the shadow radius, the deviation occurs when $M$ is close to the value of $L_*$. The time-like deflection also produces a higher value of  $\hat{\alpha}$, and the lower the impact parameter, the greater the deflection. Note that in this plot, $b = 10M$ is still in the regime for weak deflection angle since this is higher than the critical impact parameter $b_\text{crit} = 3\sqrt{3}\mathcal{M}$. We will also use this information for the weak deflection for micro-black holes and strong deflection angle.
\begin{figure} [!ht]
    \includegraphics[width=0.49\textwidth]{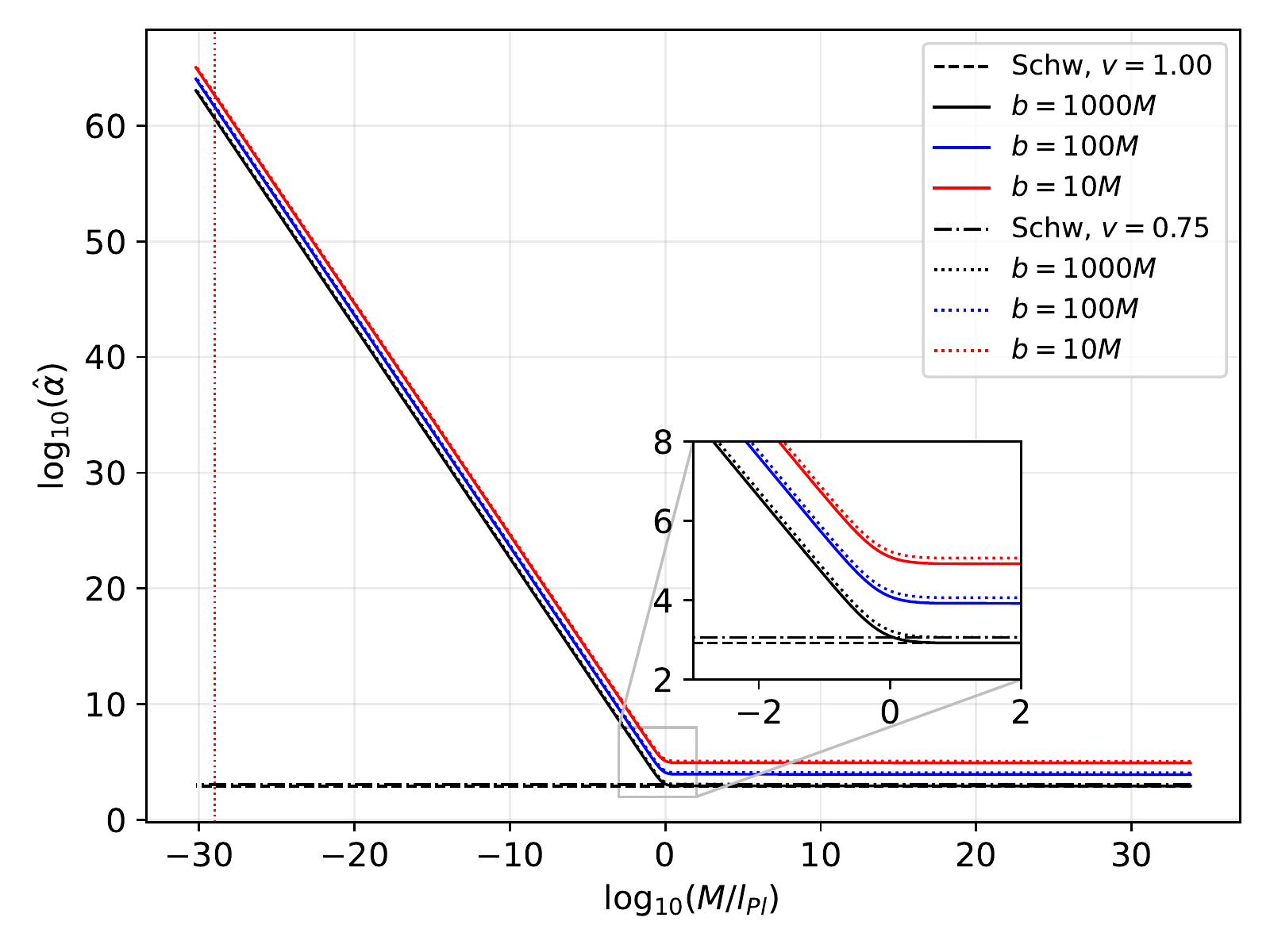}
    \caption{Weak deflection angle by micro-black hole. The red vertical dotted line corresponds to the mass of the neutrino.}
    \label{wdaelem}
\end{figure}
As a final note with our plots, showing how $\hat{\alpha}$ changes as the mass $M$ varies has its shortcomings since $b/M$ is constant. For instance, if we consider the mass of the Earth, $b = 1000M$ is equal to $4.4$ m, which is too small in comparison to the radius of the Earth ($6371$ km). Thus, the line plot in Fig. \ref{wdabh} may have its range of validity relative to the chosen value of the impact parameter. Such a result has a critical implication as far as the GEUP model in this study is concerned. One can verify that if the dimensional reduction is used in the metric in Eq. \eqref{eqn:3} to calculate $\hat{\alpha}$, that is when $l_\text{Pl} = 1$ \cite{Mureika:2018gxl}, one can observe a very high value for $\hat{\alpha}$ for low mass compact objects (such as Earth for example).

We also applied the weak deflection angle for micro-black holes \cite{Calmet:2014dea,Hawking:1971ei}. We do this by plotting $\hat{\alpha}$ vs. $\log_{10}(M/l_\text{Pl})$ in Fig. \ref{wdaelem}. Note that when one geometrized the Planck mass, the Planck length is obtained, and for simplicity, we plotted $\hat{\alpha}$ in terms of $M/l_\text{Pl}$. Qualitatively, we observe the same features as the weak deflection for astrophysical black holes. Here, we see that the deviation begins to manifest if the $\log_{10}(M/l_\text{Pl}) \sim 0$, and these are the masses that is comparable to $l_\text{Pl}$ ($\sim 2.176\text{x}10^{-8}$ kg in metric units). In this case, $\hat{\alpha} \sim 114815 \mu$as and can be detectable if one directs a photon at an impact parameter of $b \sim 1.62\text{x}10^{-32}$ m. Such a particle is still massive, and its physical dimension may cause a collision instead of a deflection. Weak deflection may occur unless we compress the particle to allow such a small value for the impact parameter. In the plot, the vertical dotted line is the neutrino's mass. We can see that $\hat{\alpha} \sim 3.89\text{x}10^{60} \mu$as for $b = 1000M$. Such a large weak deflection angle can be made smaller by increasing $b$. However, the main problem in this case is that we cannot observe neutrinos at rest.

\section{Strong Deflection Angle} \label{sec4}
Near the black hole region, specifically in the critical impact parameter, the deflection angle is described by the strong deflection expression as shown in the papers \cite{bozza2002gravitational, tsukamoto2017deflection, zhao2017strong}. The photonsphere region is crucial in strong deflection calculation; hence, we use Eq. \eqref{erph}. Following the mentioned seminal papers, we obtain the strong deflection angle as
\begin{equation}
    \label{6}
    \hat{\alpha}_\text{str} = -\bar{a} \ln\left(\frac{b_{0}}{b_\text{crit}}-1\right)+\bar{b}+\mathcal{O}(b-b_\text{crit}),
\end{equation}
where the $\bar{a}$ and $\bar{b}$ are coefficients of strong deflection. Here, $b_{0}$ and $b_\text{crit}$ corresponds to the impact parameter evaluated at the closest approach and critical impact parameter, respectively. The coefficients of the strong deflection are calculated using the equations \cite{tsukamoto2017deflection},
\begin{equation}
\label{eqn:7}
\bar{a} = \sqrt{\frac{2B(r_\text{ph})C(r_\text{ph})}{C''(r_\text{ph})A(r_\text{ph})-A''(r_\text{ph})C(r_\text{ph})}},    
\end{equation}
and
\begin{equation}
    \label{eqn:8}
    \bar{b} = \bar{a}\ln \Bigg[r_\text{ph}\left( \frac{C''(r_\text{ph})}{C(r_\text{ph})} - \frac{A''(r_\text{ph})}{A(r_\text{ph})} \right)\Bigg]+I_\text{R}(r_\text{ph})-\pi,
\end{equation}
where $A(r_\text{ph})$, $B(r_\text{ph})$, and $C(r_\text{ph})$ are metric functions evaluated at the photon sphere region. The double prime signifies second derivative with respect to $r$ evaluated at the photonsphere, $r \rightarrow r_\text{ph}$.

\begin{figure*}
    \centering
    \includegraphics[width=0.49\textwidth]{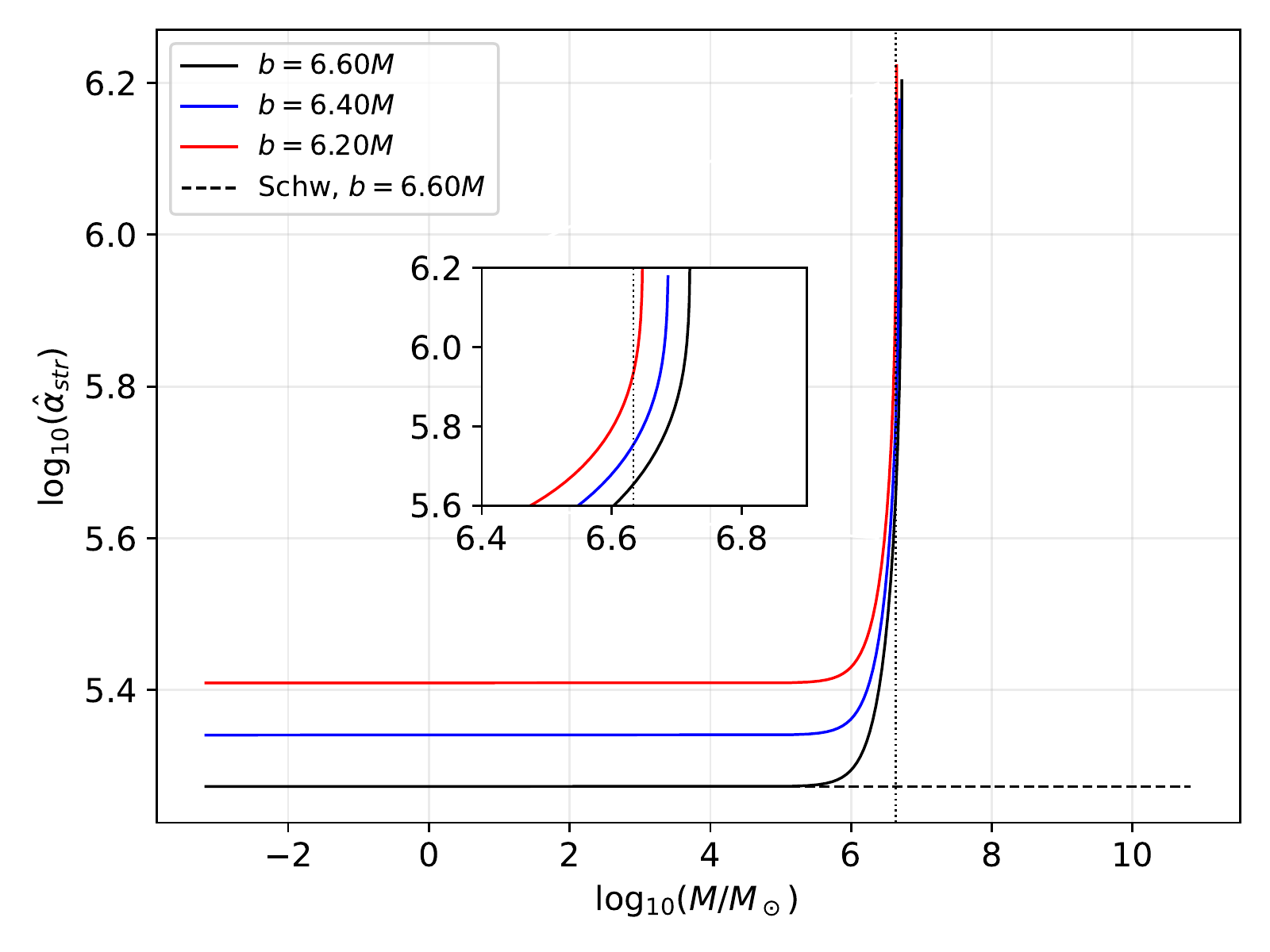}
    \includegraphics[width=0.49\textwidth]{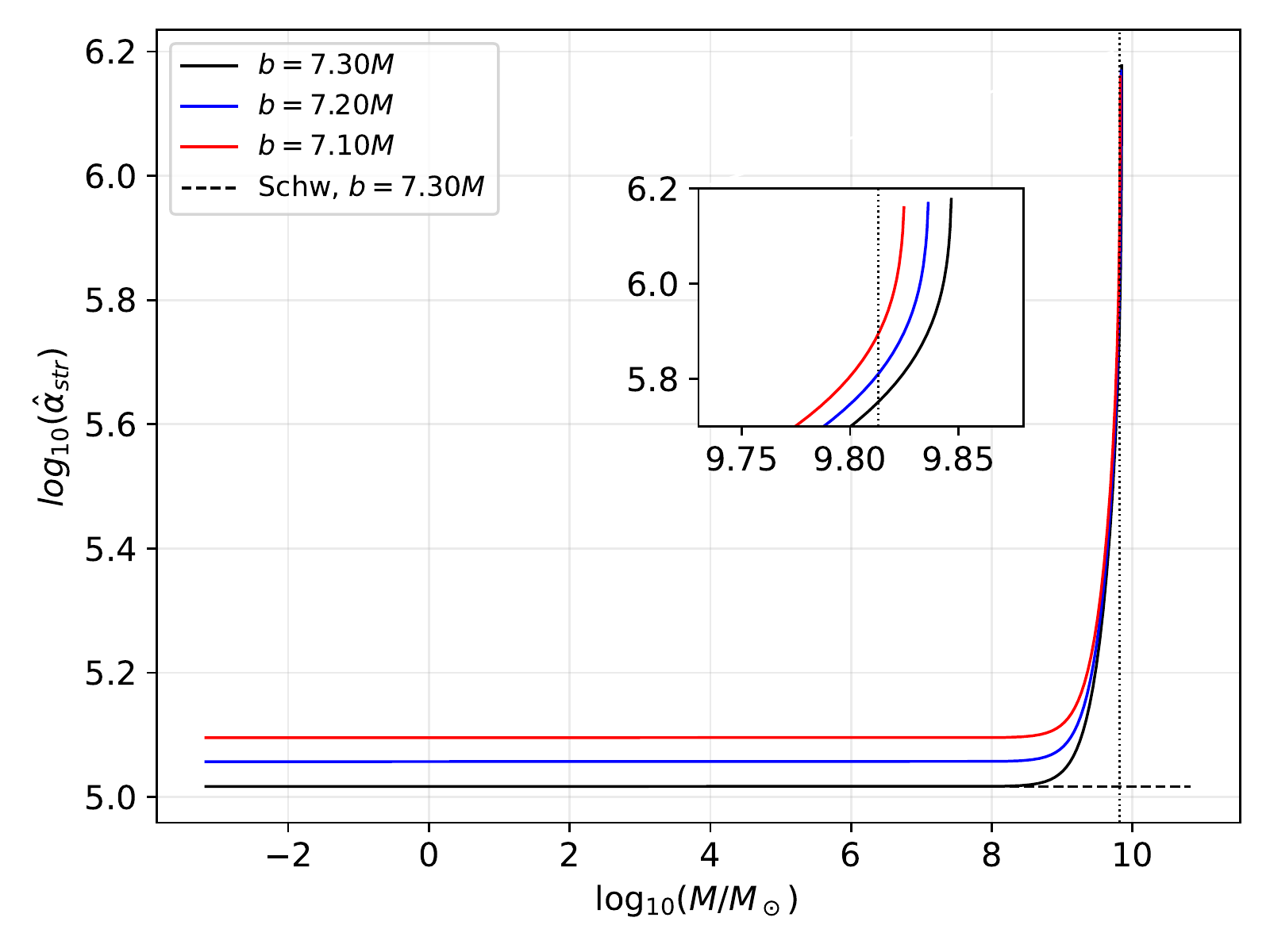}
    \caption{Behavior of strong deflection angle by Sgr. A* (left) and M87* (right). The black vertical dotted line is the corresponding mass of the SMBH.}
    \label{SDAplot}
\end{figure*}

\begin{figure}
    \includegraphics[width=0.49\textwidth]{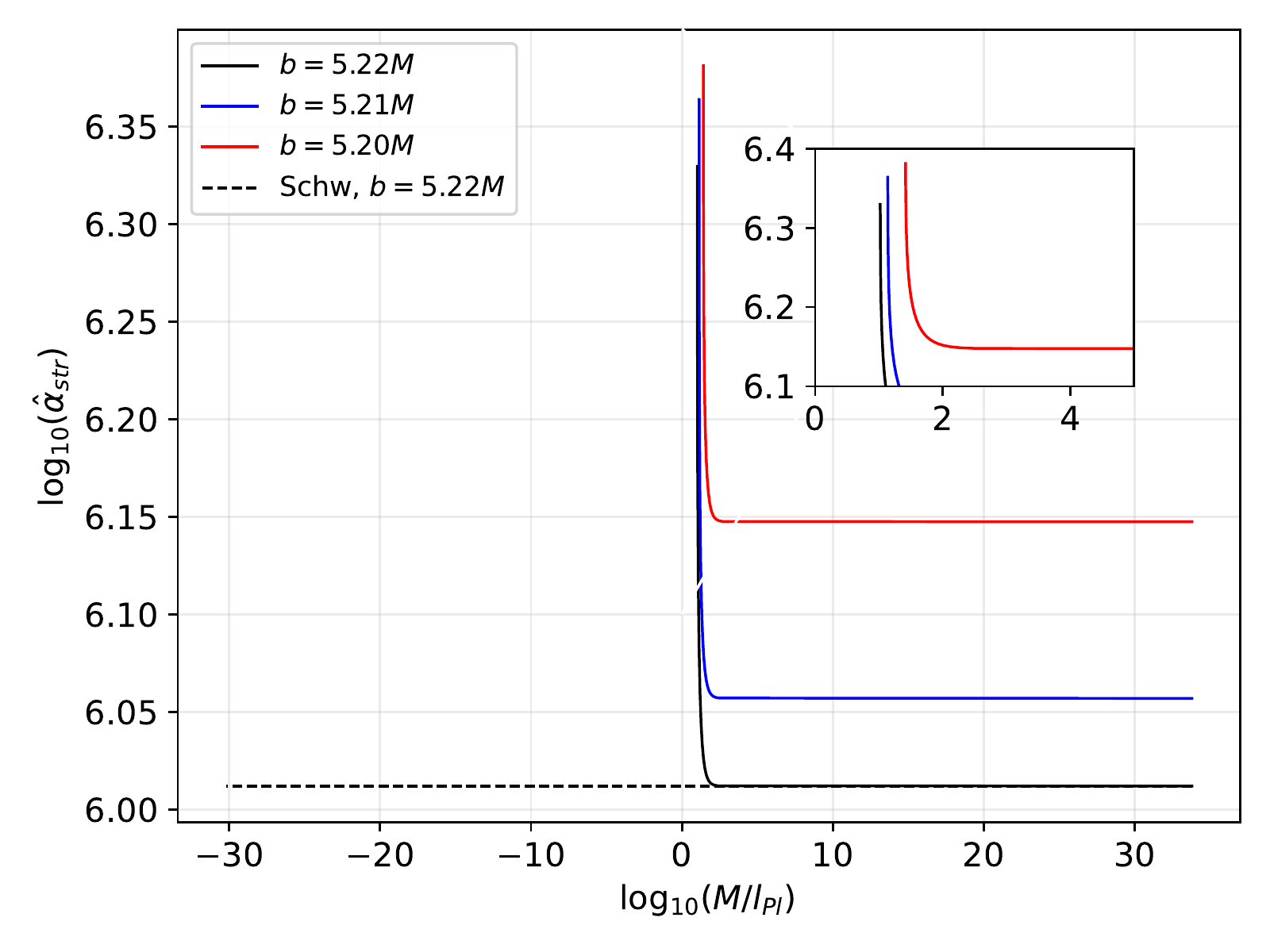}
    \caption{Strong deflection angle by micro-black holes.}
    \label{sdaelem}
\end{figure}

The second term in Eq. \eqref{eqn:8} can be calculated using the procedure illustrated in \cite{zhao2017strong, tsukamoto2017deflection} where, 
\begin{equation}
    \label{eqn:9}
    I_\text{R}(r_\text{ph}) = \int^{1}_{0} \Bigg[\frac{2(1-A_\text{ph})\sqrt{A(z, r_\text{ph})B(z, r_\text{ph})}}{A'(z, r_\text{ph})C(z, r_\text{ph}) \sqrt{\frac{A_\text{ph}}{C_\text{ph}}-\frac{A(z, r_\text{ph})}{C(z, r_\text{ph})}}} \Bigg]dz, 
\end{equation}
and $A(z, r_\text{ph})$, $B(z, r_\text{ph})$, and $C(z, r_\text{ph})$ are metric functions $A(r)$, $B(r)$, and $C(r)$ evaluated using the new variable \cite{tsukamoto2017deflection},
\begin{equation}
\label{eqn:10}
    z \equiv 1 - \frac{r_\text{ph}}{r}.
\end{equation}
We express Eq. \eqref{eqn:10} in terms of $r$ and substitute it to the metric functions. Applying the expression in Eqs. \eqref{eqn:7}-\eqref{eqn:9} to the black hole metric, we find
\begin{equation}
    \label{eqn:11}
    \bar{a} = 1,
\end{equation}
and
\begin{equation}
\label{eqn:12}
    \bar{b} = \ln\left[216(7-4\sqrt{3})\right] - \pi.
\end{equation}
When the $\alpha$, and $\beta$ are set to zero, it retrieves the Schwarzschild expression for strong deflection. Then we get the final expression as \cite{Bisnovatyi-Kogan:2008lga}, 
\begin{equation}
    \label{eqn:13}
    \hat{\alpha}_\text{str} = -\ln\left[\frac{b}{b_\text{crit}}-1\right] - 0.40023,
\end{equation}
where the critical impact parameter is expressed in Eq. \eqref{ebcrit} \cite{Pantig:2022ely}, resulting to Eq. \eqref{ebcrit2}. In choosing the value of $b$ it is essential to note that the ratio, $b/b_\text{crit}$, must not be significantly far from $1$. At $b_\text{crit} = b$ Eq. \eqref{eqn:13} diverges. It shows that the photonsphere captures particles in this region. In the plots below, we have chosen $b$ (in units of $M$) such that it is slightly greater than $b_\text{crit} = 3\sqrt{3}\mathcal{M}$. We have successfully plotted the strong deflection angle showing how GEUP affects astrophysical and micro-black holes. See Figs. \ref{SDAplot} and \ref{sdaelem}. In order to avoid extremely high values in these plots, we used a logarithmic scale.
In Fig. \ref{SDAplot}, the strong deflection angle curves are steeper than the weak deflection angle. While we observe the same feature that the low impact parameter produces a higher deflection angle, we see that the deviations due to the GEUP in the strong deflection regime occur early (at lower mass) than the weak deflection angle, thus giving an enhanced detectability.

Due to Eq. \eqref{eqn:13}, there is some value for mass $M$ where the strong deflection ceases, and this is near the value of the GEUP parameters $L_*$ and $l_\text{Pl}$ (see also Fig. \ref{sdaelem}). Without the influence of GEUP, the strong deflection angle seems to have no limit for any values of mass $M$ (as shown by the dashed black line). The same feature can be observed for micro-black holes. Again, while strong deflection is theoretically possible for small particles, a problem in its detectability is looming in the impact parameters since $b$ might be so small compared to the particle's physical dimension.

\section{Conclusion} \label{conclu}
While the effects of GUP and EUP are commonly analyzed separately in the literature, our study in this paper is about unifying these two quantum corrections as applied to black hole physics. Motivated by the paper in Ref. \cite{Mureika:2018gxl}, we investigated the effect of GEUP on the shadow and lensing for astrophysical black holes and very small particles viewed as quantum black holes.

We first find constraints to the values of $L_*$ using astrophysical data from the EHT. For $2\sigma$ level of uncertainty, we found an upper bound $L_* \sim 2.985\text{x}10^{10}$ m for Sgr. A*, while we found $L_* \sim 3.264\text{x}10^{13}$ m for M87*. Interestingly for M87*, there is a value for $L_*$ which crosses the mean of the shadow diameter, which is $L_* \sim 7.950\text{x}10^{13}$ m. We note that this order of magnitudes agrees with the constraints of gravitational lensing observables, position, magnification, and differential time delays in \cite{Lu:2019wfi}. We also examined how the shadow radius behaves based on the position of the observer from the GEUP black hole in this work. Results indicate that a BH with GEUP generally follows the same pattern for the shadow radius curve in the Schwarzschild case. In particular, the GEUP parameter $L_*$ generally increases the shadow radius for black hole masses with the same order of magnitude as $L_*$. We also did not find any influence of GUP on the shadow of astrophysical black holes. Shadow for quantum black holes is also investigated. Here, as the particle's mass $M$ under GEUP correction decreases, we see that its corresponding shadow increases. The position of detectors also affects the radius of such shadows. Lastly, it has been shown that $L_*$ does not affect the micro-black hole's shadow.

Alternatively, we probe more into the effects of GEUP by considering the strong and weak deflection angles. For the weak deflection angle, the main result indicates that deviation caused by GEUP occurs when the masses are comparable to the fundamental length scales. Such a deviation occurs early in strong deflection angle. Furthermore, due to the fundamental length scales, there is a limitation for the occurrence of strong deflection angle. For example, if we hypothetically observe the deflection angle by a neutrino, strong deflection cannot be applied due to the limitations imposed by $l_\text{Pl}$.

Nonetheless, the weak deflection angle is still a better probe since it can be applied for very high impact parameters. The drawback is that measurement may not be possible due to the quantum nature of a particle. Finally, as far as the GEUP model in this study is concerned, the strong and weak deflection angle cannot probe whether $L_*$ affects the quantum realm, and vice versa. Lowering the value of $L_*$ may give an interesting result, but it may have some implications in the astrophysical phenomena that might be ruled out by observation. In theory, this direction might be worth investigating.

\bibliography{ref}
\bibliographystyle{apsrev}

\end{document}